\documentclass{bmvc2k}
\usepackage{multirow}
\usepackage{multicol}
\usepackage{booktabs}
\usepackage{amsmath, amssymb}
\usepackage{type1cm}
\usepackage{subfigure}
\usepackage{comment}

\title{Acoustic-based 3D Human Pose Estimation Robust to Human Position}

\runninghead{OUMI et al.}{Acoustic-based 3D human pose estimation}

\addauthor{Yusuke Oumi}{youmi@keio.jp}{1}
\addauthor{Yuto Shibata}{yuto19990715@gmail.com}{1}
\addauthor{Go Irie}{goirie@ieee.org}{1,2}
\addauthor{Akisato Kimura}{akisato@ieee.org}{3}
\addauthor{Yoshimitu Aoki}{aoki@elec.keio.ac.jp}{1}
\addauthor{Mariko Isogawa}{mariko.isogawa@keio.jp}{1,4}

\addinstitution{
 Keio University, Japan
}
\addinstitution{
 Tokyo University of science, Japan
}
\addinstitution{
 Nippon Telegraph and\\ Telephone Corporation, Japan
}
\addinstitution{
 JST Presto, Japan
}

\def\eg{\emph{e.g}\bmvaOneDot}

\def\etal{\emph{et al}\bmvaOneDot}

\begin{document}

\maketitle

\begin{abstract}
This paper explores the problem of 3D human pose estimation from only low-level acoustic signals. 
The existing active acoustic sensing-based approach for 3D human pose estimation implicitly assumes that the target user is positioned along a line between loudspeakers and a microphone. Because reflection and diffraction of sound by the human body cause subtle acoustic signal changes compared to sound obstruction, the existing model degrades its accuracy significantly when subjects deviate from this line, limiting its practicality in real-world scenarios.
To overcome this limitation, we propose a novel method composed of a position discriminator and 
reverberation-resistant model.
The former predicts the standing positions of subjects and applies adversarial learning to extract subject position-invariant features. 
The latter utilizes acoustic signals before the estimation target time as references to enhance robustness against the variations in sound arrival times due to diffraction and reflection. We construct an acoustic pose estimation dataset that covers diverse human locations and demonstrate through experiments that our proposed method outperforms existing approaches.
\end{abstract}

\begin{figure*}
\begin{center}
\includegraphics[width=.8\linewidth]{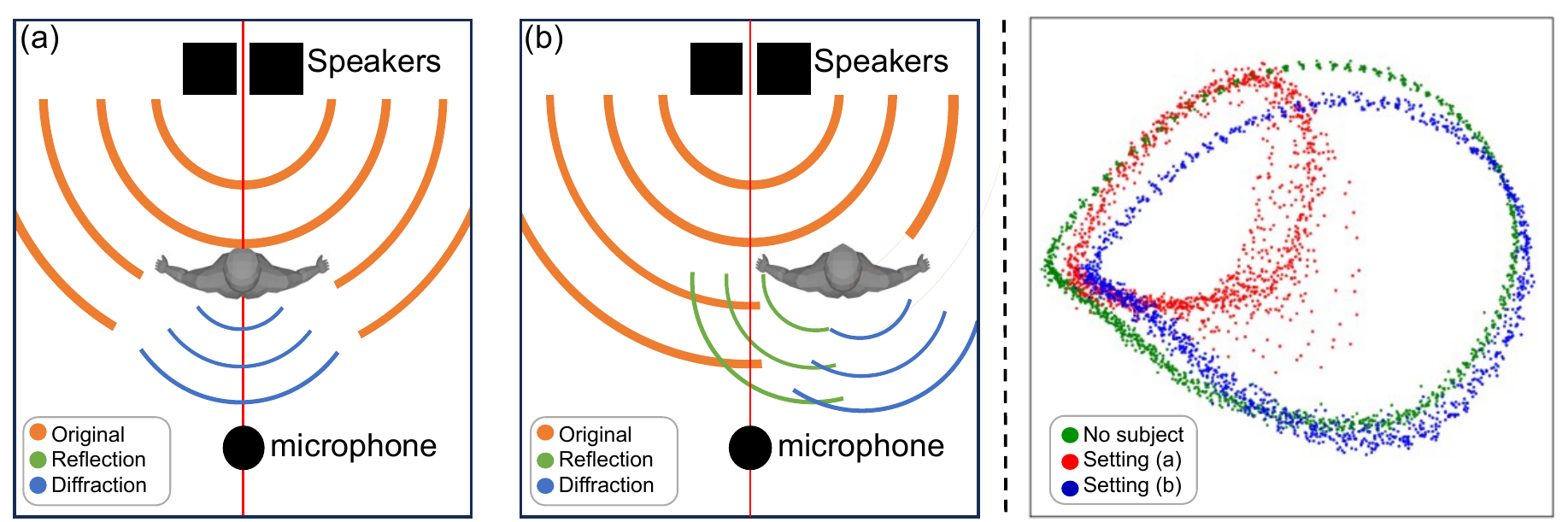}{}
\end{center}
\vspace{-5mm}
   \caption{Left: acoustic-signal based human pose estimation with different target subject positions. Unlike the existing method that (a) utilizes obscured acoustic signals by the target subject positioned along the line between the microphone and loudspeaker, we aim to estimate (b) poses with the subject who is away from this line, which is the challenging task due to the presence of signal reflection and diffraction. Right: principal component analysis of acoustic features depending on a subject's standing position.}
   \vspace{-5mm}
\label{fig:fig1}
\end{figure*}

\section{Introduction}
\label{sec:intro}

Human pose estimation has diverse applications including rehabilitation support, elderly monitoring, and disaster relief efforts. Traditional approaches to 3D human pose estimation have primarily employed RGB videos and images \cite{simple_yet_martinez_2017_ICCV,volumetric_pred_Pavlakos_2017_CVPR}, transient light \cite{nlos_pose}, event data \cite{eventlifting_scarpellini_2021,efficient_chen_2022}, radio frequency (RF)/Wi-Fi signals \cite{RFpose3D_zhao_2018_sigcomm,wipose}, and millimeter wave \cite{m3track,mmmesh_xue_2021}. 
Additionally, methods that combine some of these approaches as a multimodal framework also exist~\cite{mri_An_2022, Mm-fi_yang_2024}.
However, optical signals face challenges such as obstruction and poor performance in low-light conditions \cite{low-light_lee_2023}. Furthermore, RGB video and images acquire high-resolution measured data, which raises concerns regarding the protection of personal information. 
Wireless signal-based methods are restricted in environments employing precision machinery, such as medical facilities or aircraft.

One possible solution to these challenges is the utilization of acoustic signals. Acoustic signals have much longer wavelengths (meter scale) compared to optical signals (nanometer scale) or RF/Wi-Fi signals (centimeter scale). Therefore, acoustic signals are more susceptible to diffraction and less affected by obstruction. Moreover, acoustic signals offer consistent performance irrespective of lighting conditions and their usage is not hindered by the presence of precision machinery.

Recent studies have explored passive acoustic sensing for gesture recognition and human pose estimation by leveraging human speech ~\cite{audio2gestures_li_2021_ICCV,speech2pose}, ambient sounds ~\cite{listentolook_gao_2020_CVPR}, or the sound of playing a musical instrument ~\cite{audio2body}. These methods require sounds produced by the subjects themselves, which limits the use case.
Alternatively, Shibata \etal proposed a 3D human pose estimation approach using active acoustic sensing with Time-Stretched-Pulse (TSP) signals \cite{sound2pose}. In this approach, a subject is positioned between a speaker and microphone (see Fig. \ref{fig:fig1}(a)), where the speaker repeatedly emits TSP signals to create an acoustic field, and human poses are estimated based on how the acoustic field distorts as a subject moves.
However, this method primarily relies on how the acoustic signal emitted from the speaker is obstructed by the human body to estimate the human pose. 
It implicitly assumes that the target subject is positioned on a straight line between the speaker and the microphone, although in the real world, meeting such constraints is extremely rare.
Through the preliminary experiments, we found that the estimation accuracy significantly decreases when the subject deviates from this line, due to the difficulty of capturing subtle changes in sound signals caused by human body movements.
Fig. \ref{fig:fig1}(c) 
visualizes the acoustic features used as input to the model. The dimensions of these features are reduced by the Principal Component Analysis (PCA). The acoustic features in the settings without any subject and those shown in figures (a) and (b) are represented in different colors. From this figure, it is confirmed that the acoustic features when a person moves away from this line (blue dots) approach the features when there is no subject (green dots), indicating sound diffraction and reflection convey much less human pose information than sound obstruction caused by a person standing on the aforementioned line (red dots).

To overcome this limitation, this paper proposes an acoustic-based 3D human pose estimation method, which remains effective regardless of the subject's standing positions.
While Shibata \etal primarily relied on signal obstruction by the human body as their main clue, in this paper, we also consider cases where the position of the person is not on the straight line connecting the speaker and the microphone, as shown in Fig. \ref{fig:fig1}(b). 
Therefore, it is necessary to consider signal diffraction and reflection from the subject as well as signal obstruction.
From a technical perspective, this implies the need to solve two extremely challenging issues: (i) The relatively long wavelengths of acoustic signals tend to cause specular reflections off the surface of the human body. Consequently, the sound intensity of the reflected acoustic signals is greatly influenced by the positions of reflection and recording microphones. (ii) The arrival time of sound emitted from a speaker until it is recorded can vary due to signal diffraction and reflection.

In this paper, we aim to develop methods capable of addressing these challenges. First, to enhance robustness against variations in the subject's position, we introduce a position discriminator module. 
This module uses intermediate features of the pose estimation module to predict human positions, while the pose estimation module is trained to maximize the uncertainty of human positions, through adversarial training.
Furthermore, to achieve robust pose estimation against changes in the arrival time of sound due to sound diffraction and reflection,
we propose to introduce a reference window into the pose estimation module to consider signals prior to the target time to be estimated.
Additionally, we perform data augmentation by shifting the phase of the acoustic signal, which allows for a reduction in the amount of data per subject location, enabling the preparation of a dataset that covers diverse positions. As the first attempt at non-invasive 3D human pose estimation regardless of the subject's position, we construct a new dataset containing data from positions away from the straight line connecting the speaker and the microphones. 

In summary, the technical contributions of this study are as follows: 
(1) We have worked towards realizing a practical non-invasive 3D human pose estimation method based on active acoustic signals while subjects are placed in multiple positions. 
(2) We introduced a position discriminator module to enhance robustness against variations in the subject's standing position. Additionally, we constructed a pose estimation model that considers acoustic signals prior to the estimation target time to achieve robust estimation against changes in sound arrival times due to signal diffraction and reflection. (3) To effectively learn from limited data, we performed data augmentation by shifting the phase of the acoustic signal. (4) As the first attempt to estimate non-invasively 3D human pose regardless of the subject's position, we constructed a dataset containing data from multiple positions away from the straight line connecting the speaker and the microphones.

\section{Related Work}
\label{sec:related}

\noindent
{\bf{Human Pose Estimation with Different Modalities.}}
Human pose estimation is a traditional task in the field of computer vision and is expected to be used for a wide range of applications. 
Many studies have utilized RGB-based methods \cite{openpose_tpami,simple_yet_martinez_2017_ICCV,volumetric_pred_Pavlakos_2017_CVPR}, which allow for relatively easy pose estimation through cameras.
However, these methods face decreased accuracy in low-light (\eg, a dark room, night road) or occluded environments. Additionally, their ability to capture extensive information can lead to significant privacy concerns.
Event-based methods \cite{eventlifting_scarpellini_2021,efficient_chen_2022} are also influenced by occlusion, affecting estimation accuracy.

In response to these challenges, methods using other modalities, such as RF/Wi-Fi signals  \cite{RF-pose,RFpose3D_zhao_2018_sigcomm,wipose,gopose} and millimeter waves \cite{m3track,mmmesh_xue_2021}, have also been studied.
While these methods can perform estimations independent of lighting conditions, they are limited in environments with sensitive electronic equipment where the use of wireless signals is restricted, and their performance can be hindered by obstructions such as water and metal \cite{RF-pose}.
To overcome these limitations, our research focuses on the utilization of acoustic sensing for human pose estimation.
\\
\noindent
{\bf{Acoustic Sensing Related to Human Activities.}}
Our research aligns closely with fields that utilize human speech and musical instrument sounds for estimating gestures, including joint positions, through passive acoustic sensing \cite{speech2pose, audio2body, audio2gestures_li_2021_ICCV,listentolook_gao_2020_CVPR}. 
To estimate human pose, this type of research relies on sounds such as speech audio, which includes semantics, making it more prone to the potential identification of personal information.
Moreover, there is a study that requires subjects to wear devices that emit sounds to estimate gestures \cite{audiotouch}. However, this method is limited by the necessity for subjects to wear such devices. Thus, we propose non-invasive active acoustic sensing for pose estimation, enabling broader application of our method across various scenarios.
\\
\noindent
{\bf{Acoustic Sensing-based Scene Estimation.}}
Methods for estimating room environment using acoustic signals often employ geometric transformations based on the Room Impulse Response (RIR) to predict reflection locations~\cite{pose_kernel_lifter,acoustic_reconsruction,neural_acoustic_fields}. In these techniques, the room environment is treated as a system with sounds emitted by a speaker as the input and sounds captured by a microphone as the output to obtain RIR. 
Accurate capture of the echoes in relation to the sounds emitted by the speaker is necessary for these geometric transformations. 
However, in this paper, we need to keep sending the echo to estimate the dynamic poses.
Consequently, the overlapping of residual sounds from previous instances with newly emitted sounds complicates the precise calculation of RIR. Therefore, following Shibata \etal\cite{sound2pose}, we avoid geometric reflection position estimation and instead transmit Time Stretched Pulse (TSP) signals from the speaker. We then generate acoustic feature vectors from the signals received by the microphone and employ machine learning to attempt pose estimation which is robust to the target position.

\vspace{-2mm}
\section{Methodology}
\label{sec:method}

\begin{figure*}
\begin{center}
\includegraphics[width=.95\linewidth]{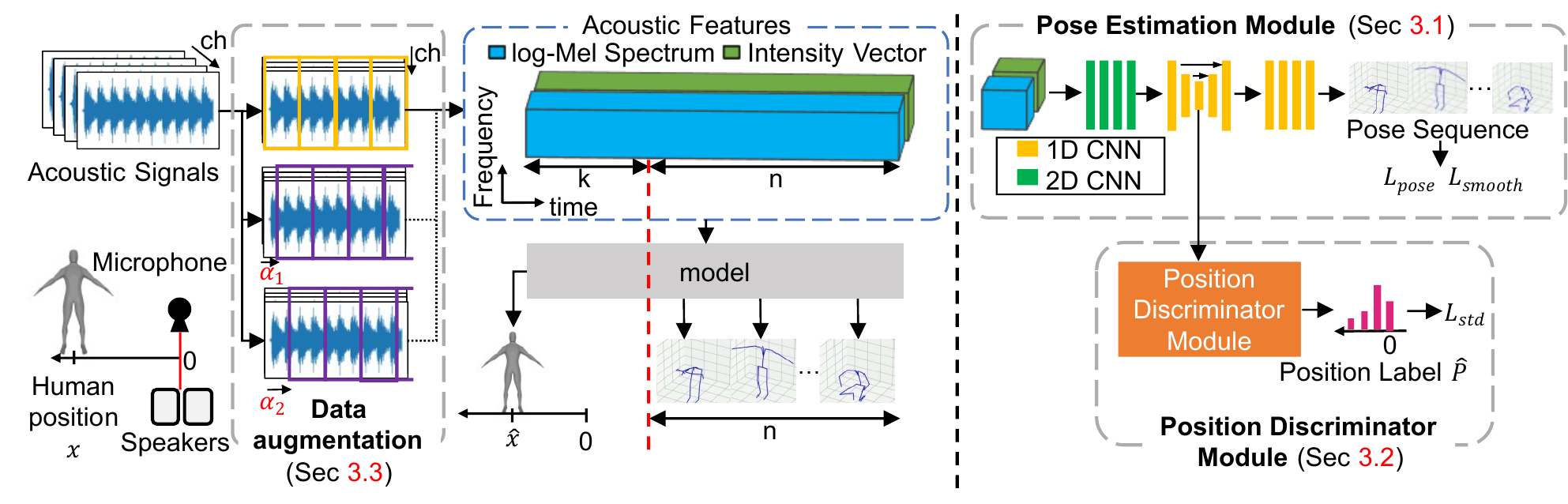}{}
\end{center}
\vspace{-7mm}
\caption{Proposed Framework.}
\vspace{-5mm}
\label{fig:proposed_method}
\end{figure*}

Our goal is to estimate a sequence of 3D human poses $\mathbf{p}=[p_1,p_2,...,p_T]$ from an acoustic signal sequence $\mathbf{s}=[s_1,s_2,...,s_T]$, segmented into fixed lengths from audio signals recorded by a microphone. $T$ is the sequence length, and $s_t$ and $p_t$ refer to the $t$-th elements of the acoustic signal sequence and the 3D pose sequence, respectively. 
Following~\cite{sound2pose}, we use TSP signal, a periodic signal whose frequency varies over time for active acoustic sensing.
It replaces the pose estimation using the recorded signals for TSP signals with the analysis of room impulse responses utilizing the spatial reverberation characteristics.
The acoustic signal sequence $\mathbf{s}$ is recorded in B-Format using a 4-channel ambisonics microphone.

The proposed framework shown in Figure \ref{fig:proposed_method} consists of three components: the acoustic feature generation module, which converts the acoustic signal $\mathbf{s}$ into an acoustic feature vector $\mathbf{a}=[a_1,a_2,...,a_T]$, the pose estimation module $f$, and the position discriminator module, which determines the standing position of the subject.
Following Shibata \etal \cite{sound2pose}, acoustic feature $\mathbf{a}$ consists of two components: the log-Mel Spectrum $I^{logmel} \in \mathbb{R}^{b \times 4}$ and the Intensity Vector~\cite{intensity_vector} which is represented as $I^{intensity} \in \mathbb{R}^{b \times 3}$. 
Each element $a_t$ of $\mathbf{a}$ is in the form of $b\times 7$ tensor, where $b$ represents the number of Mel filter banks. 
In the following subsections, we will discuss the pose estimation module (Sec \ref{sec:31}), the position discriminator module (Sec \ref{sec:32}), and data augmentation (Sec \ref{sec:33}).

\subsection{Pose Estimation Module}
\label{sec:31}
The pose estimation module $f(\mathbf{a})$ consists of 2D convolutional layers and 1D convolutional layers. 
The function $f$ simultaneously estimates the $n$ consecutive poses $[p_i, p_{i+1}, ..., p_{i+n-1}]$. During this process, the corresponding acoustic features $[a_i, a_{i+1},..., a_{i+n-1}]$ are influenced by reverberant acoustic signals from several frames earlier, delayed due to reflection and diffraction. 
Therefore, the proposed method considers the time series relationships of sound by including acoustic information from $k$ frames prior to the target sequence, thus utilizing $n+k$ frames of acoustic features $[a_{i-k}, a_{i-k+1}, ..., a_{i+n-1}]$ as input for the pose estimation.
With the variable $\theta$ that contains all trainable parameters and weight hyperparameters $w_{\alpha}$, $w_{\beta}$, and $w_{\gamma}$, the training objective is to minimize the following loss function $\mathcal{L}$.
    \vspace{-2mm}
\begin{eqnarray}
    \mathcal{L} = w_{\alpha} \mathcal{L}_{pose} + w_{\beta} \mathcal{L}_{smooth} + w_{\gamma} \mathcal{L}_{std}
\label{loss}
\end{eqnarray}
    \vspace{-4mm}

\noindent
The loss function $\mathcal{L}_{pose}$ related to the human pose is calculated as the Mean Squared Error (MSE) between $i$-th ground truth pose $p_i$ and predicted pose $\hat{p}_i$.
The loss function $\mathcal{L}_{smooth}$ is used to smoothly connect consecutive poses  $\hat{p}_i$ and $\hat{p}_{i-1}$.
\vspace{-2mm}
\begin{eqnarray}
    \mathcal{L}_{pose} (\theta)=\frac{1}{T} \sum_{i=1}^{T} \|\hat{p}_{i} - p_{i}\|_2 
\end{eqnarray}
\vspace{-8mm}
\begin{eqnarray}
   \mathcal{L}_{smooth} (\theta)=\frac{1}{T-1} \sum_{i=2}^{T} \|(\hat{p}_{i} - \hat{p}_{i-1}) - (p_{i} - p_{i-1})\|_2 
\end{eqnarray}
\vspace{-4mm}

\noindent
$\mathcal{L}_{std}$ is the loss function used for adversarial learning with the position discriminator module. Details are provided in Sec \ref{sec:32}.

\subsection{Position Discriminator Module}
\label{sec:32}
The position discriminator module is composed of a single fully connected layer and uses the intermediate outputs from the pose estimation module as inputs to learn the subject's position. 
The pose estimation module engages in adversarial learning against the position discriminator module to extract features that are independent of position, enhancing the robustness of human positions.

Here, one of the most straightforward ways of implementation for position estimation within the position discriminator module is to utilize regression.
However, introducing regression-based predictors into adversarial learning is known to potentially cause gradient explosions. To address this issue, we treat the distance from the line connecting the speaker and microphone as a label and train position discrimination based on classification rather than regression.
Therefore, continuous locations of the subject are represented using soft labels (linear combinations). 
The position discriminator module 
outputs the label value $\hat{P}$ for the standing position. 
To enable accurate pose estimation for any standing position by the position estimation module, we utilize the loss function $\mathcal{L}_{std}$ to generate feature representations that are invariant to the human position.
    \vspace{-3mm}
\begin{eqnarray}
    \mathcal{L}_{std}= \frac{1}{T'} \sum_{j=1}^{T'}\mathrm{STD}(\hat{P}_j)
\end{eqnarray}
Here, $\mathrm{STD}(\cdot)$ denotes the standard deviation.
$\hat{P}_j$ is the output of the position discriminator module, which corresponds to one of the 
$n$ poses $[p_i,p_{i+1},...,p_{i+n-1}]$.
Accordingly, $T'$  means the number of acoustic and pose sequences and is given by $T' = T/n$.
When the position discriminator module successfully estimates the subject's position, one of the label values in $\hat{P}$ will have a significantly higher value, while the others will have smaller values. Conversely, if the position estimation fails, the label values in $\hat{P}$ will exhibit similar magnitudes. From this observation, the standard deviation of $\hat{P}$, denoted as $\mathrm{STD}(\hat{P})$, increases as the certainty of the position estimation improves. 
Consequently, $L_{std}$ will increase in value as the accuracy of the position estimation increases.

\subsection{Data Augmentation}
\label{sec:33}

A general challenge in deep learning-based human pose estimation tasks is the need for a large amount of training data. As we tackle a new task, we cannot leverage existing large-scale datasets. Also, this paper assumes that subjects are positioned in multiple positions, necessitating data collection for each position. Therefore, compared to existing work that assumes subjects are standing in fixed positions, our data collection cost becomes significantly higher, making the collection of large amounts of real-world data quite costly.
Therefore, we also propose to introduce data augmentation for this task.
By shifting the starting time of one period of the TSP signal by $\alpha$ time units (equivalent to shifting the phase of the acoustic signal), and generating acoustic features from the shifted received signal, we perform data augmentation. The ground truth poses are similarly shifted by the time parameter $\alpha$, and the average pose associated with the acoustic signals used to create acoustic features is determined.

\begin{figure*}
\begin{center}
\includegraphics[width=.85\linewidth]{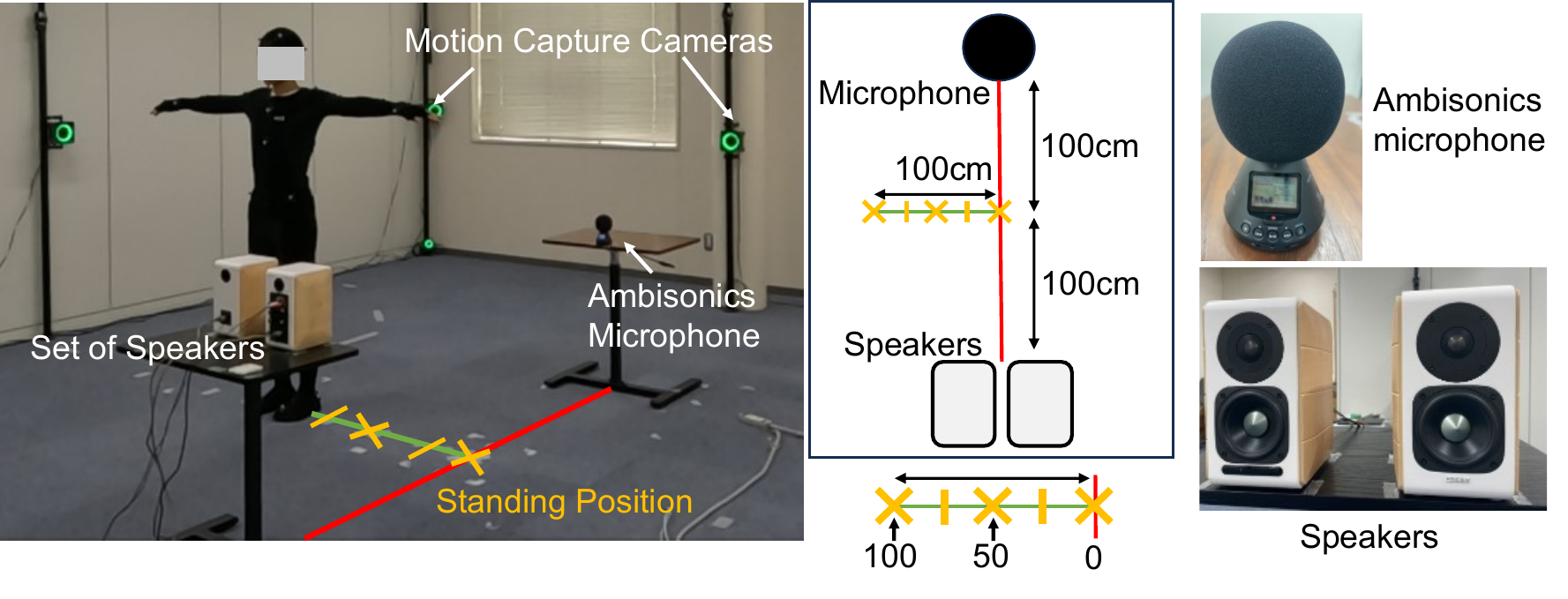}{}
\end{center}
\vspace{-8mm}
\caption{Experimental setup and equipment}
\vspace{-5mm}
\label{fig:fig3}
\end{figure*}

\section{Experimental Settings}
\label{sec:setting}
\subsection{Dataset and Setup}
\label{sec:41}
For active acoustic sensing, we utilized a pair of loudspeakers (Edifier ED-S880DB) and the Ambisonics microphone (Zoom H3-VR). To obtain ground truth 3D pose, we employed the motion capture system (OptiTrack) with 16 cameras (see Fig.~\ref{fig:fig3}). The experiments were conducted in a classroom environment with background noise and reverberation.

Five male subjects were asked to stand at five positions along the line connecting the speaker and the microphone: directly on the line, and at 25 cm, 50 cm, 75 cm, and 100 cm away from this line. They were asked to take various poses including walking, squatting, bowing, standing, T-pose, and intermediate poses between these movements. We used 21 joints including the head, neck, both shoulders, both arms, both forearms, both hands, waist, both thighs, both shins, both feet, both toes, hip, and spine. The dataset size was approximately 3.5 hours in total.
\vspace{-3mm}

\subsection{Baselines}
\label{sec:42}
We compared our method against the following three methods:
(1) Jiang \etal \cite{wipose} as one of the state-of-the-art methods for based 3D human pose estimation with low-dimensional input signals like our method. Specifically, the original method utilizes Wi-Fi signals and introduces an LSTM network. Since the original method is Wi-Fi-based, we modified the input layer of this method so that it can use our log-Mel Spectrum and Intensity Vector as input.
(2) Ginosar \etal's method~\cite{speech2pose} that estimates human gestures from speech sounds. This method employs a CNN-based network with temporal convolutions, using only log-Mel Spectrum as the input acoustic feature.
(3) Shibata \etal \cite{sound2pose}, which is the most relevant method to ours, estimates the pose of subjects located along a straight line between a speaker and a microphone in the form of active acoustic sensing. This method also employs a CNN-based network that processes temporal information through temporal convolutions like (2), and it utilizes both log-Mel Spectrum and Intensity Vector as inputs.
\vspace{-3mm}

\subsection{Evaluation Metrics}
\label{sec:43}

In this paper, three types of evaluation metrics were used: Root Mean Squared Error (RMSE), Mean Absolute Error (MAE), and Percentage of Correct Keypoints (PCK). RMSE and MAE are metrics calculated from the true poses and the estimated poses. PCK calculates the proportion of correctly estimated keypoints compared to the true keypoints, considering distances below a certain threshold as correct. In this study, we use PCKh@0.5, where h represents the distance between the keypoints of the head and neck, and the threshold is set to half of this distance.
\vspace{-3mm}

\subsection{Implementation Details}
\label{sec:44}
For all methods, we used Adam \cite{Adam} as optimizer.
Ginosar \etal and Shibata \etal's methods simultaneously estimate 12 frame poses using 12 frames of acoustic features as inputs. In contrast, the proposed method estimates 8 frame poses simultaneously from 24 frames of acoustic features as described in Sec~\ref{sec:31}~($n=8,k=16$).
For the loss calculation in Eq. \ref{loss}, weight parameters were set as $w_{\alpha}=w_{\gamma}=1, w_{\beta}=10$. 
Furthermore, for the parameter $\alpha$ for the data augmentation, one-third and two-thirds of the size of each acoustic signal sequence element were used. 

\section{Experimental Results}
\label{sec:result}

\subsection{Comparison with Other Baselines}
\label{sec:51}
In this paper, four out of the five subjects were used as training data to train the model $f$, and the fifth subject's data, not included in training, was used for testing. This process was repeated for each subject to calculate the average estimation accuracy for five subjects. The table \ref{tab:baseline} shows a qualitative comparison with the baseline method.
The proposed method outperforms all others across three evaluation metrics. 

\begin{figure*}
\begin{center}
\vspace{-5mm}
\includegraphics[width=.85\linewidth]{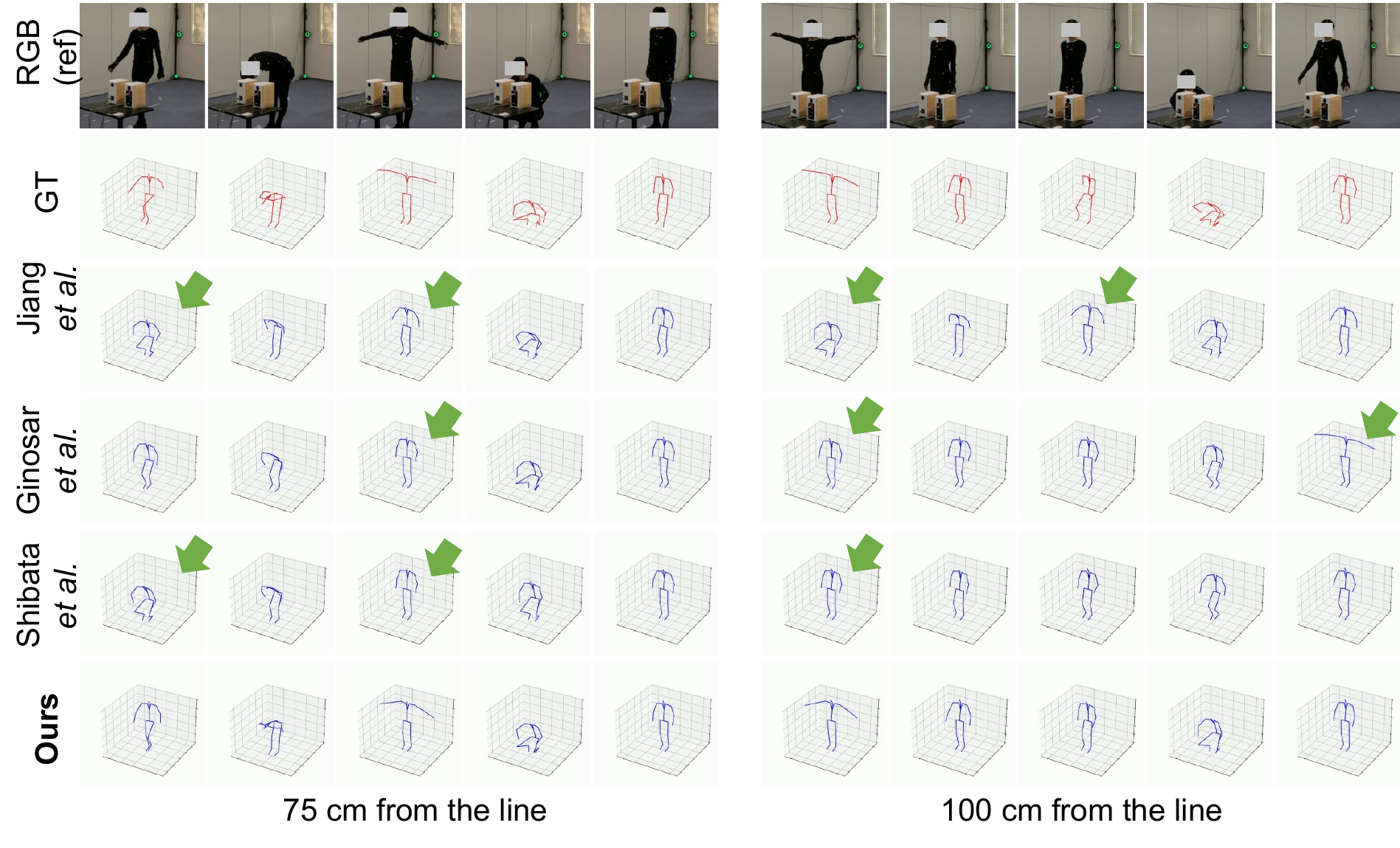}{}
\end{center}
\vspace{-7mm}
\caption{The qualitative results of 75 cm and 100 cm from the line}
\label{fig:qualitative}
\end{figure*}

Figure \ref{fig:qualitative} shows the qualitative comparison. 
To distinguish between the ``T-pose'' and ``standing'', it is necessary to detect the arms raised horizontally, as the positions of the torso and legs remain the same. This requires discerning subtle differences in the sounds reflected off the arms, which is a more delicate task compared to other movements. The proposed method effectively captures these subtle acoustic differences, resulting in more accurate T pose estimations compared to baseline methods.
Additionally, the methods by Ginosar \etal and Shibata \etal frequently misestimate poses in the first half of pose sequences.
The method by Jiang \etal does not utilize temporal convolution operations, which results in unstable pose predictions.

\begin{table}[t]
    \begin{minipage}[b]{0.5\linewidth}
    \centering
        \caption{Comparison against baselines}
        \vspace{-5mm}
        \begin{center}\resizebox{0.95\linewidth}{!}{ 
        \begin{tabular}[t]{@{\hskip 1mm}lcccc@{\hskip -1mm}ccc@{\hskip 1mm}ccc}
        \toprule
        \multirow{3}{*}{Method} & 
        \multirow{2}{*}{RMSE} & \multirow{2}{*}{MAE} & \multirow{2}{*}{\shortstack[c]{PCKh \\ @0.5}}\\
        \\
        & ($\downarrow$) & ($\downarrow$) & ($\uparrow$)\\ 
        \midrule
        Jiang \etal ~\cite{wipose} & 0.75 & 0.40 & 0.48\\
        Ginosar \etal ~\cite{speech2pose} & 0.65 & 0.33 & 0.55 \\
        Shibata \etal ~\cite{sound2pose} & 0.66 & 0.35 & 0.53\\
        Ours & \textbf{0.53} & \textbf{0.28} & \textbf{0.60}\\
        \bottomrule
        \label{tab:baseline}
        \end{tabular}
    \label{tab:my_label}
        }    
        \end{center}

    \end{minipage}
    \hfill
    \begin{minipage}[b]{0.5\linewidth}
            \centering
            \caption{Ablation Study}
            \vspace{-5mm}
        \begin{center}\resizebox{0.88\linewidth}{!}{ 
        \begin{tabular}[t]{@{\hskip 1mm}lcccc@{\hskip -1mm}ccc@{\hskip 1mm}ccc}
        \toprule
        \multirow{3}{*}{Method} & 
        \multirow{2}{*}{RMSE} & \multirow{2}{*}{MAE} & \multirow{2}{*}{\shortstack[c]{PCKh \\ @0.5}}\\
        \\
        & ($\downarrow$) & ($\downarrow$) & ($\uparrow$)\\ 
        \midrule
        Ours w/o Adv & 0.55 & 0.29 & 0.56\\
        Ours w/o Prior & 0.69 & 0.35 & 0.55 \\
        Ours w/o Aug & 0.58 & 0.31 & 0.55\\
        Ours & \textbf{0.53} & \textbf{0.28} & \textbf{0.60}\\
        \bottomrule
        \label{tab:ablation}
        \end{tabular}
    \label{tab:my_label}
    }
    \end{center}
    \end{minipage}
\end{table}

\subsection{Ablative Analysis}
The Ablation study was conducted to individually evaluate the effects of the three technical contributions introduced in the proposed method.
Table \ref{tab:ablation} shows a quantitative comparison excluding each component one at a time.
In this table, adversarial learning with the position discrimination module is denoted as "Adv", the proposed method that uses information prior to the target time is referred to as "Prior", and data augmentation using phase shifts is represented as "Aug".
The results indicate that the complete the proposed method achieves the best results across all three evaluation metrics. Particularly, the inclusion of information prior to estimated time was found to dominate the accuracy improvements. 
Detailed comparisons on the estimation's precision using previous time information are provided in Section \ref{sec:53}, and discussions on the second most contributing factor, data augmentation through phase shifting, are in Section \ref{sec:54}.

\subsection{Comparison by Input Size}
\label{sec:53}
In our method, 24 samples of acoustic features are used as input to the model, which then outputs the pose corresponding to the last 8 samples of these 24. We tested reduced input sizes of 8 and 16 samples and an increased size of 32 samples.
Table \ref{tab:input} shows the quantitative evaluation for different input sizes. Input size of 8 samples resulted in lower accuracy across all metrics.
When the input size is reduced to 16, the PCK is the same as the proposed method, and the rough behavior is relatively well estimated.
However, the lack of information prior to the estimation target time particularly lowered the RMSE values.
Conversely, increasing the input to 32 samples also resulted in decreased accuracy. This configuration involves using inputs that reach 1.2 seconds back from the target estimated time, which exceeds the typical reverberation time in a classroom environment. Therefore, it is likely that the model overfits acoustical information that is less relevant to the actual poses.

\begin{table}[t]
    \begin{minipage}[b]{0.45\linewidth}
    \centering
        \caption{Effect of model input sizes}
        \vspace{-2mm}
        \begin{center}\resizebox{0.85\linewidth}{!}{ 
        \begin{tabular}{@{\hskip 1mm}lcccc}
        \toprule
        \multirow{3}{*}{\shortstack[c]{Input\\Size}} & 
        \multirow{2}{*}{RMSE} & \multirow{2}{*}{MAE} & \multirow{2}{*}{\shortstack[c]{PCKh \\ @0.5}}\\
        \\
        & ($\downarrow$) & ($\downarrow$) & ($\uparrow$)\\ 
        \midrule
        8 (w/o Prior)& 0.69 & 0.35 & 0.55\\
        16 & 0.58 & 0.29 & \textbf{0.60} \\
        24 & \textbf{0.53} & \textbf{0.28} & \textbf{0.60}\\
        32 & 0.59 & 0.31 & 0.56\\
        \bottomrule
        \label{tab:input}
        \end{tabular}
    \label{tab:my_label}
        }    
        \end{center}

    \end{minipage}
    \hfill
    \begin{minipage}[b]{0.45\linewidth}
            \centering
            \caption{Effect of data augmentation}
            \vspace{-2mm}
        \begin{center}\resizebox{0.78\linewidth}{!}{ 
        \begin{tabular}{@{\hskip 1mm}lcccc}
        \toprule
        \multirow{3}{*}{\shortstack[c]{Number \\ of Data}} & 
        \multirow{2}{*}{RMSE} & \multirow{2}{*}{MAE} & \multirow{2}{*}{\shortstack[c]{PCKh \\ @0.5}}\\
        \\
        & ($\downarrow$) & ($\downarrow$) & ($\uparrow$)\\ 
        \midrule
        w/o Aug & 0.58 & 0.31 & 0.55\\
        double & 0.54 & \textbf{0.28} & 0.57 \\
        triple & \textbf{0.53} & \textbf{0.28} & \textbf{0.60}\\
        quadruple & \textbf{0.53} & \textbf{0.28} & \textbf{0.60}\\
        \bottomrule
        \label{tab:aug}
        \end{tabular}
    \label{tab:my_label}
    }
    \end{center}
    \end{minipage}
    \vspace{-5mm}
\end{table}

\subsection{Comparison by Data Augmentation}
\label{sec:54}

In the proposed method, data augmentation was performed by tripling the number of training frames using phase shift hyperparameter $\alpha$. We also evaluated the effects of doubling and quadrupling the number of training data frames. Table \ref{tab:aug} shows the quantitative evaluation when varying the amount of training data, highlighting that our augmentation method contributes to accuracy improvements in all metrics. However, we can see that the effects of data augmentation appear to saturate beyond three times the original data. This is likely because while slight phase changes increase the diversity of acoustic features, such slight time delays have little impact on the distribution of target pose sequences.

\subsection{Evaluation with In Plain Clothes Dataset}
\label{sec:cloth}

\begin{figure}[t]
\begin{center}
\includegraphics[width=0.85\textwidth]{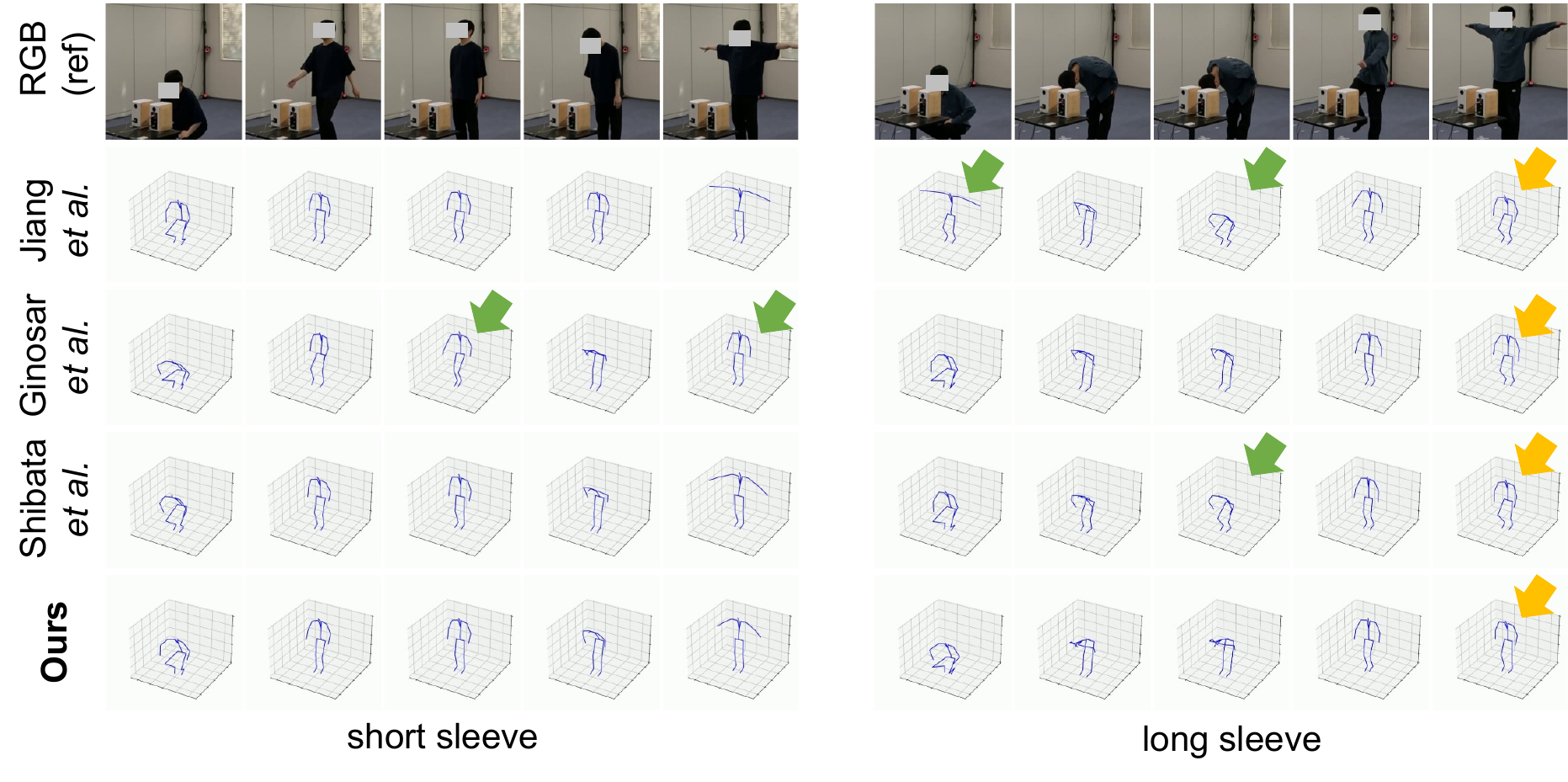}
\end{center}
\vspace{-5mm}
\caption{Comparison by Plain Clothes Dataset}
\vspace{-3mm}
\label{fig:shifuku}
\end{figure}

To assess estimation accuracy in real-world settings, we conducted a qualitative evaluation with subjects wearing casual clothing. 
Fig. \ref{fig:shifuku} shows the qualitative comparison results of two subjects who wore short- and long-sleeved clothing, different from the motion capture suits subjects wore during the training data collection.
Green and yellow arrows indicate poses where the estimation failed.
As discussed in Sec. \ref{sec:51}, estimating the T pose becomes particularly challenging when subjects are positioned away from the line. 
For the subject in short sleeves, the shape of the arms is similar to that when wearing a Mocap suit, resulting in minimal degradation in T pose estimation accuracy. 
However, for the subject wearing long sleeves, the acoustic reflection characteristics of the arms differ from those in a body-fitting Mocap suit. Consequently, a decrease in the accuracy of T pose estimation has been observed (see yellow arrows).
\vspace{-3mm}

\section{Conclusion}
\label{sec:conclusion}
In this paper, we addressed the challenge of estimating human poses at positions away from the line connecting the speaker and microphone. 
We introduce adversarial learning for position estimation, sequence size determination based on prior time-step information, and phase-shift data augmentation. These approaches allowed us to achieve superior accuracy across all evaluation metrics compared to baseline methods.

However, our method has some limitations.  We are unable to estimate the poses at unseen subject's positions not included in the training. 
This limitation arises because our model learns the echo characteristics of locations present within the training dataset. Consequently, when a subject moves to an unseen position, the echo characteristics differ from those the model has learned, hindering accurate pose estimation.
Additionally, the data used in this paper were all collected in a single classroom. The acoustic signals captured by the microphones can vary depending on the size of the room and the reflectivity of the surfaces. It will be necessary in future work to address the variations in estimation accuracy caused by such environmental characteristics.

In future studies, we aim to improve our method to effectively estimate poses for subjects moving across a broader range of settings, including those at unseen positions, and various room settings.

\bibliography{myref}

\begin{thebibliography}{26}
\providecommand{\natexlab}[1]{#1}
\providecommand{\url}[1]{\texttt{#1}}
\expandafter\ifx\csname urlstyle\endcsname\relax
  \providecommand{\doi}[1]{doi: #1}\else
  \providecommand{\doi}{doi: \begingroup \urlstyle{rm}\Url}\fi

\bibitem[An et~al.(2022)An, Li, and Ogras]{mri_An_2022}
Sizhe An, Yin Li, and Umit Ogras.
\newblock mri: Multi-modal 3d human pose estimation dataset using mmwave, rgb-d, and inertial sensors.
\newblock \emph{{Advances in Neural Information Processing Systems (NeurIPS)}}, 35:\penalty0 27414--27426, 2022.

\bibitem[Cao et~al.(2019)Cao, Iqbal, Kong, Galindo, Wang, and Plumbley]{intensity_vector}
Yin Cao, Turab Iqbal, Qiuqiang Kong, M~Galindo, Wenwu Wang, and Mark~D Plumbley.
\newblock Two-stage sound event localization and detection using intensity vector and generalized cross-correlation.
\newblock In \emph{Proc. Detection Classification Acoustic Scenes Events (DCASE) Challange}, 2019.

\bibitem[Cao et~al.(2021)Cao, Hidalgo, Simon, Wei, and Sheikh]{openpose_tpami}
Zhe Cao, Gines Hidalgo, Tomas Simon, Shih-En Wei, and Yaser Sheikh.
\newblock {OpenPose}: Realtime multi-person {2D} pose estimation using part affinity fields.
\newblock \emph{IEEE Transactions on Pattern Analysis and Machine Intelligence (TPAMI)}, 43\penalty0 (1):\penalty0 172--186, 2021.

\bibitem[Chen et~al.(2022)Chen, Shi, Ye, Yang, Sun, and Wang]{efficient_chen_2022}
Jiaan Chen, Hao Shi, Yaozu Ye, Kailun Yang, Lei Sun, and Kaiwei Wang.
\newblock Efficient human pose estimation via 3d event point cloud.
\newblock In \emph{International Conference on 3D Vision (3DV)}, pages 1--10, 2022.

\bibitem[Gao et~al.(2020)Gao, Oh, Grauman, and Torresani]{listentolook_gao_2020_CVPR}
Ruohan Gao, Tae-Hyun Oh, Kristen Grauman, and Lorenzo Torresani.
\newblock Listen to look: Action recognition by previewing audio.
\newblock In \emph{IEEE/CVF Conference on Computer Vision and Pattern Recognition (CVPR)}, pages 10457--10467, 2020.

\bibitem[Ginosar et~al.(2019)Ginosar, Bar, Kohavi, Chan, Owens, and Malik]{speech2pose}
Shiry Ginosar, Amir Bar, Gefen Kohavi, Caroline Chan, Andrew Owens, and Jitendra Malik.
\newblock Learning individual styles of conversational gesture.
\newblock \emph{IEEE/CVF Conference on Computer Vision and Pattern Recognition (CVPR)}, pages 3497--3506, 2019.

\bibitem[Isogawa et~al.(2020)Isogawa, Yuan, O'Toole, and Kitani]{nlos_pose}
Mariko Isogawa, Ye~Yuan, Matthew O'Toole, and Kris~M Kitani.
\newblock Optical non-line-of-sight physics-based 3d human pose estimation.
\newblock In \emph{IEEE/CVF Conference on Computer Vision and Pattern Recognition (CVPR)}, pages 7013--7022, 2020.

\bibitem[Jiang et~al.(2020)Jiang, Xue, Miao, Wang, Lin, Tian, Murali, Hu, Sun, and Su]{wipose}
Wenjun Jiang, Hongfei Xue, Chenglin Miao, Shiyang Wang, Sen Lin, Chong Tian, Srinivasan Murali, Haochen Hu, Zhi Sun, and Lu~Su.
\newblock Towards 3d human pose construction using wifi.
\newblock \emph{International Conference on Mobile Computing and Networking (MobiCom)}, pages 1--14, 2020.

\bibitem[Kingma and Ba(2015)]{Adam}
Diederik~P. Kingma and Jimmy Ba.
\newblock Adam: A method for stochastic optimization.
\newblock In \emph{International Conference on Learning Representations (ICLR)}, 2015.

\bibitem[Kong et~al.(2022)Kong, Xu, Yu, Chen, Ma, Chen, Chen, and Kong]{m3track}
Hao Kong, Xiangyu Xu, Jiadi Yu, Qilin Chen, Chenguang Ma, Yingying Chen, Yi-Chao Chen, and Linghe Kong.
\newblock m3track: mmwave-based multi-user 3d posture tracking.
\newblock In \emph{Proceedings of the 20th Annual International Conference on Mobile Systems, Applications and Services (MobiSys)}, pages 491--503, 2022.

\bibitem[Kubo et~al.(2019)Kubo, Koguchi, Shizuki, Takahashi, and Hilliges]{audiotouch}
Yuki Kubo, Yuto Koguchi, Buntarou Shizuki, Shin Takahashi, and Otmar Hilliges.
\newblock Audiotouch: Minimally invasive sensing of micro-gestures via active bio-acoustic sensing.
\newblock In \emph{Proceedings of the 21st international conference on human-computer interaction with mobile devices and services (MobileHCI)}, pages 1--13, 2019.

\bibitem[Lee et~al.(2023)Lee, Rim, Jeong, Kim, Woo, Lee, Cho, and Kwak]{low-light_lee_2023}
Sohyun Lee, Jaesung Rim, Boseung Jeong, Geonu Kim, Byungju Woo, Haechan Lee, Sunghyun Cho, and Suha Kwak.
\newblock Human pose estimation in extremely low-light conditions.
\newblock In \emph{IEEE/CVF Conference on Computer Vision and Pattern Recognition (CVPR)}, pages 704--714, 2023.

\bibitem[Li et~al.(2021)Li, Kang, Pei, Zhe, Zhang, He, and Bao]{audio2gestures_li_2021_ICCV}
Jing Li, Di~Kang, Wenjie Pei, Xuefei Zhe, Ying Zhang, Zhenyu He, and Linchao Bao.
\newblock Audio2gestures: Generating diverse gestures from speech audio with conditional variational autoencoders.
\newblock In \emph{IEEE/CVF International Conference on Computer Vision (ICCV)}, pages 11293--11302, 2021.

\bibitem[Luo et~al.(2022)Luo, Du, Tarr, Tenenbaum, Torralba, and Gan]{neural_acoustic_fields}
Andrew Luo, Yilun Du, Michael Tarr, Josh Tenenbaum, Antonio Torralba, and Chuang Gan.
\newblock Learning neural acoustic fields.
\newblock \emph{{Advances in Neural Information Processing Systems (NeurIPS)}}, 35:\penalty0 3165--3177, 2022.

\bibitem[Martinez et~al.(2017)Martinez, Hossain, Romero, and Little]{simple_yet_martinez_2017_ICCV}
Julieta Martinez, Rayat Hossain, Javier Romero, and James~J Little.
\newblock A simple yet effective baseline for 3d human pose estimation.
\newblock In \emph{IEEE/CVF International Conference on Computer Vision (ICCV)}, pages 2640--2649, 2017.

\bibitem[Pavlakos et~al.(2017)Pavlakos, Zhou, Derpanis, and Daniilidis]{volumetric_pred_Pavlakos_2017_CVPR}
Georgios Pavlakos, Xiaowei Zhou, Konstantinos~G. Derpanis, and Kostas Daniilidis.
\newblock Coarse-to-fine volumetric prediction for single-image 3d human pose.
\newblock In \emph{IEEE/CVF Conference on Computer Vision and Pattern Recognition (CVPR)}, July 2017.

\bibitem[Ren et~al.(2022)Ren, Wang, Wang, Tan, Chen, and Yang]{gopose}
Yili Ren, Zi~Wang, Yichao Wang, Sheng Tan, Yingying Chen, and Jie Yang.
\newblock Gopose: 3d human pose estimation using wifi.
\newblock \emph{Proceedings of the ACM on Interactive, Mobile, Wearable and Ubiquitous Technologies (IMWUT)}, 6\penalty0 (2):\penalty0 1--25, 2022.

\bibitem[Scarpellini et~al.(2021)Scarpellini, Morerio, and Del~Bue]{eventlifting_scarpellini_2021}
Gianluca Scarpellini, Pietro Morerio, and Alessio Del~Bue.
\newblock Lifting monocular events to 3d human poses.
\newblock In \emph{IEEE/CVF Conference on Computer Vision and Pattern Recognition (CVPR) Workshop}, pages 1358--1368, 2021.

\bibitem[Shibata et~al.(2023)Shibata, Kawashima, Isogawa, Irie, Kimura, and Aoki]{sound2pose}
Yuto Shibata, Yutaka Kawashima, Mariko Isogawa, Go~Irie, Akisato Kimura, and Yoshimitsu Aoki.
\newblock Listening human behavior: 3d human pose estimation with acoustic signals.
\newblock In \emph{IEEE/CVF Conference on Computer Vision and Pattern Recognition (CVPR)}, pages 13323--13332, 2023.

\bibitem[Shlizerman et~al.(2018)Shlizerman, Dery, Schoen, and Kemelmacher-Shlizerman]{audio2body}
Eli Shlizerman, Lucio Dery, Hayden Schoen, and Ira Kemelmacher-Shlizerman.
\newblock Audio to body dynamics.
\newblock In \emph{IEEE/CVF Conference on Computer Vision and Pattern Recognition (CVPR)}, pages 7574--7583, 2018.

\bibitem[Straubinger et~al.(2022)Straubinger, Xiao, and Rhodin]{acoustic_reconsruction}
Tim Straubinger, Robert Xiao, and Helge Rhodin.
\newblock Learned acoustic reconstruction using synthetic aperture focusing.
\newblock In \emph{IEEE International Conference on Acoustics, Speech and Signal Processing (ICASSP)}, pages 1606--1610, 2022.

\bibitem[Xue et~al.(2021)Xue, Ju, Miao, Wang, Wang, Zhang, and Su]{mmmesh_xue_2021}
Hongfei Xue, Yan Ju, Chenglin Miao, Yijiang Wang, Shiyang Wang, Aidong Zhang, and Lu~Su.
\newblock mmmesh: Towards 3d real-time dynamic human mesh construction using millimeter-wave.
\newblock In \emph{Proceedings of the 20th Annual International Conference on Mobile Systems, Applications and Services (MobiSys)}, pages 269--282, 2021.

\bibitem[Yang et~al.(2024)Yang, Huang, Zhou, Chen, Xu, Yuan, Zou, Lu, and Xie]{Mm-fi_yang_2024}
Jianfei Yang, He~Huang, Yunjiao Zhou, Xinyan Chen, Yuecong Xu, Shenghai Yuan, Han Zou, Chris~Xiaoxuan Lu, and Lihua Xie.
\newblock Mm-fi: Multi-modal non-intrusive 4d human dataset for versatile wireless sensing.
\newblock \emph{{Advances in Neural Information Processing Systems (NeurIPS)}}, 2024.

\bibitem[Yang et~al.(2022)Yang, Fan, Isler, and Park]{pose_kernel_lifter}
Zhijian Yang, Xiaoran Fan, Volkan Isler, and Hyun~Soo Park.
\newblock Posekernellifter: Metric lifting of 3d human pose using sound.
\newblock In \emph{IEEE/CVF Conference on Computer Vision and Pattern Recognition (CVPR)}, pages 13179--13189, 2022.

\bibitem[Zhao et~al.(2018{\natexlab{a}})Zhao, Li, Abu~Alsheikh, Tian, Zhao, Torralba, and Katabi]{RF-pose}
Mingmin Zhao, Tianhong Li, Mohammad Abu~Alsheikh, Yonglong Tian, Hang Zhao, Antonio Torralba, and Dina Katabi.
\newblock Through-wall human pose estimation using radio signals.
\newblock In \emph{IEEE/CVF Conference on Computer Vision and Pattern Recognition (CVPR)}, pages 7356--7365, 2018{\natexlab{a}}.

\bibitem[Zhao et~al.(2018{\natexlab{b}})Zhao, Tian, Zhao, Alsheikh, Li, Hristov, Kabelac, Katabi, and Torralba]{RFpose3D_zhao_2018_sigcomm}
Mingmin Zhao, Yonglong Tian, Hang Zhao, Mohammad~Abu Alsheikh, Tianhong Li, Rumen Hristov, Zachary Kabelac, Dina Katabi, and Antonio Torralba.
\newblock Rf-based 3d skeletons.
\newblock In \emph{Proceedings of the 2018 Conference of the ACM Special Interest Group on Data Communication (SIGCOMM)}, pages 267--281, 2018{\natexlab{b}}.

\end{thebibliography}
\end{document}